# Chemical order lifetimes in liquids in the energy landscape paradigm.

Luz-Maria Martinez and C. Austen Angell

*Department of Chemistry and Biochemistry, Arizona State University*

*Tempe, Arizona, USA  85287-1604*

**Abstract**

**Recent efforts to deal with the complexities of the liquid state, particularly those of glassforming systems, have focused on the "energy landscape" as a means of dealing with the collective variables problem [1]. The "basins of attraction" that constitute the landscape features in configuration space represent a distinct class of microstates of the system. So far only the microstates that are related to structural relaxation and viscosity have been considered in this paradigm. But most of the complex systems of importance in nature and industry are solutions, particularly solutions that are highly non-ideal in character. In these, a distinct class of fluctuations exists, the fluctuations in concentration, and their character may be of much importance to the liquid properties. The mean square amplitudes of these fluctuations relate to the chemical activity coefficients [2], and their rise and decay times may be much longer than those of the density fluctuations - from which they are statistically independent. Here we provide experimental information on the character of chemical order fluctuations in viscous liquids and on their relation to the density and enthalpy fluctuations that determine the structural relaxation time, and hence the glass transition temperature. Using a spectroscopically active chemical order probe, we identify a "chemical fictive temperature", $T_{chm}$, by analogy with the "fictive temperature" $T_f$ commonly used to denote the temperature where the structural arrest occurred during cooling a**

**glassformer. Like $T_f$, $T_{chm}$ must be the same as the real temperature for the system to be in complete equilibrium. It is possible for mobile multicomponent liquids to be permanently nonergodic, insofar as $T_{chm} > T_f = T$, which must be accommodated within the landscape paradigm. We note that, in appropriate systems, an increase in concentration of slow chemically ordering units in liquids can produce a crossover to fast ion conducting glass phenomenology.**

The "liquid problem" is greatly simplified in the case of liquids that supercool, due to the development of a large separation between the time scales on which the vibrational and configurational degrees of freedom fluctuate. This separation is seen most dramatically at the glass transition where the structural fluctuations freeze out leaving, to first approximation, only the vibrational modes to contribute to the intensive thermodynamic properties of the (now solid) system. Following initial work by Goldstein [3], advantage of this situation has been taken by Stillinger [4,5] to rigorously separate the chemical potential of a liquid into two distinct components. One is the vibrational free energy it possesses when confined to one of its unique structural states ("basin of attraction"), and the other is its structural free energy due to the position of the particular basin of attraction on the energy landscape. The first relates to kinetic energy and determines the true temperature. The second relates to its potential energy and may be characterized by its "fictive" temperature. In a state of equilibrium, the two temperatures are the same [6]. A third temperature is possible in the case of multicomponent systems, and it is this third temperature with which we are concerned in this report.

The third temperature we call the "chemical fictive temperature", designated $T_{chm}$, and it is the same as the others when the chemical order has its equilibrium value. The chemical order is subject to fluctuations and, in the absence of a strong volume dependence on concentration, the chemical order (or concentration) fluctuations are statistically independent of the density fluctuations that determine the compressibility [2,7,8]. The statistical thermodynamics of the chemical order fluctuations have been analyzed by Bhatia and coworkers [2,7]. The relation of the chemical order fluctuations to the composition dependence of the Gibbs free energy G (hence to the activity coefficients describing the departures from ideal solution behavior) is

$$S_{cc} = N(<\Delta c>^2) = Nk_BT/(\partial^2 G/\partial c^2)_{T,P,N} \qquad (1)$$

(where $S_{cc}$ is the concentration-concentration structure factor, $<\Delta c>^2$ is the mean square concentration fluctuation). The validity of this expression has been verified by neutron scattering experiments which can detect the concentration fluctuations involved [8,9]. What has not been discussed in detail is the time dependence of these fluctuations and the manner in which they relate to the time dependence of density and entropy fluctuations that have been studied so intensely in recent years [10].

That the chemical order fluctuations may have different and longer relaxation times is of course well known from the detailed study of binary liquid systems near their critical solution temperatures [11]. At the consolute composition, the clustering of components during cooling will become critical if the chemical order does not become frozen first. The freezing of pre-spinodal fluctuations near a glass transition has been observed [12], and the fact that, in non-ideal solutions, these fluctuations "freeze in" *above* $T_g$ during cooling (due to their longer relaxation times) has been documented [12 (c)]. However, these studies have all involved systems that exhibit positive deviations from ideal behavior.

Here we wish to focus attention on the chemical order resulting from the alternative departure from ideality, namely that due to negative deviations from ideal mixing. Indeed such deviations are very common in glassforming systems. After all, it is the choosing of components that chemically order optimally that is behind the art of glass-making: the chemical ordering of the components must be strong enough to depress the freezing point of a multicomponent liquid far below the ideal mixing value, without being so strong as to generate new high-melting crystals [13]. It is in the vicinity of the deep eutectics that result from optimal deviations from ideal mixing, that the glassforming compositions are found [14, 15]. The parameters of the much-studied mixed LJ system [1, 16, 18] were selected on this basis, being modeled on the Ni-P eutectic metallic glassformer [15].

When the chemical order has slow dynamics relative to the structural fluctuations that determine the viscosity, then the analysis of Stillinger [4,5], in which degrees of freedom are separated for individual study, can be extended, viz.,

$$\mu_{tot} = \mu_{vib} + \mu_{pot} + \mu_{chm}$$

and the properties of the mobile liquid that depend on vibrational and structural degrees of freedom, can be evaluated at constant (frozen) chemical order.

Our primary aims in this report are (i) to document the separation of chemical ordering time-scales from structural relaxation time scales that may arise under the right circumstances, and (ii) to characterize the chemical ordering process as an almost exponential, almost linear, almost Arrhenius process, as distinct from the normal non-exponential, non-linear, super-Arrhenius relaxation processes so well documented for structural relaxation [10]. A separate interest, which remains unelucidated in this report, is to determine whether the chemical order relaxation time is dominated by the thermodynamics of chemical ordering in the same manner as the structural relaxation time seems to be dominated by the thermodynamics of structural ordering, according to recent reports [16-19].

To conduct this study we chose a system in which the chemical order can be detected by electronic spectroscopy. In earlier studies of this type [20,21] the cobalt(II) ion has been utilized because of the clear distinction between its electronic spectra when the ion is in structural states of different coordination number. In ref. 20 the alternative coordination states were four co-ordination by chloride ions and eight coordination by nitrate ion oxygens, the latter being the high energy state populated at high temperature. In the present study we provide a solvent matrix that is rich in both chloride and hydroxyl ligands. We choose a mixture of components (see methods section) in which a Co(II) probe ion distributes itself, in a manner favorable for analysis, between octahedral (oxygen as -OH) and tetrahedral (chloride) sites, the oxygen coordination being in this case the low temperature state. However, unlike the systems of previous isothermal studies, the present solution lacks other cations of high charge intensity to compete with the doubly charged cobalt ion. At least when the probe species is dilute, the chemical order detected in this experiment explores two states only. The population is temperature sensitive, with a van't Hoff enthalpy change of 11.7 kcal/mol. This is large enough that changes in chemical order that occur when the temperature is jumped 2-5K up or down (on a time scale that is short with respect to the chemical order relaxation time) are easily detected. By monitoring the absorption intensity of the tetrahedral $CoCl_4^{2-}$ site at 690 nm (see Methods section) as a function of time after temperature jumps of 2-5K, the kinetics of the chemical order change can be determined.

In Fig. 1 the time dependence of the chemical order relaxation is shown for down jumps of 3K to the temperatures 286, 283, 280, and 277K. The lengthening of the relaxation time with decreasing temperature is obvious. In contrast to the case for structural relaxation, the relaxation is independent of the direction of the jump, i.e. the relaxation is almost linear. This is shown in Fig. 2, in which a comparison is made with classical data for the non-linear relaxation of the density [22] of a silicate glass. A non-linearity even stronger than that of Fig. 2a was recently demonstrated in the high temperature (fragile) regime of the network liquid $BeF_2$, which is a strong liquid at low temperatures, but makes a transition to fragile liquid behavior at high temperatures [23]. Thus the chemical order relaxation observed in this study is of different character from the structural relaxation. Corresponding to this difference, is the finding that the chemical order relaxes almost exponentially in time, with a "stretching exponent" $\beta \approx 0.8$, unlike the density, typically $\beta = 0.5$ [10]. That a small departure from linearity remains is suggested not only by the residual non-exponentiality but also by a weak dependence of the average relaxation time on the magnitude of the jump. The latter is shown in Fig. 3 inset, which shows that for 2K jumps the relaxation is quite linear.

The origin of these effects is to be found mainly in the fact that the chemical order relaxation time is longer than that of the structural relaxation, with the result that sources of non-exponentiality are averaged out. The difference in time scales becomes more pronounced the higher the temperature. This is shown in Fig. 3, main part, in which the results of all the temperature up jump experiments are shown along with results of viscosity measurements made on the same sample using a digital viscometer [24]. Consistent with earlier data for pure sorbitol and trehalose [25], the viscosity data, combined with the molecular liquid value, $10^{11}$ Pa.s at $T_g$, show that the solvent medium is fragile in character. However the chemical order relaxation time shows a much smaller temperature dependence, approaching simple Arrhenius behavior. This behavior contrasts with that of the other examples of probes of structural relaxation in the recent literature where the relaxation is long and exponential, but where the relaxation is "slaved" to the solvent relaxation temperature dependence [26]. Unlike these latter cases, the chemical order decouples from the structural relaxation with increasing temperature. The chemical order can evidently relax on longer time scales, without affecting the more rapid relaxation of the structure around it. What we are observing here seems to be the low temperature glassforming system analog of the classical desolvation kinetics studies of the Eigen school [27]. A T-jump result for cobalt desolvation in a molecular solvent system [28] has been temperature -scaled onto our results by placing the second order kinetics relaxation times on Fig. 3 at a temperature where our "solvent" would have the same viscosity as that of ref. 28. The somewhat slower process for Ni(II) from ref. 28 is also included in Fig. 3.

Thus, during quenching at some given rate, which means moving across Fig. 3 along an isochrone (constant time scale, shorter for faster quenches), the chemical order will freeze first at the "chemical fictive temperature" $T_{chm}$ and, at some lower temperature, the structure will freeze at the usual fictive temperature T. The gap between the two will be larger the faster the quench. Conditions for the observation of "chemical aging" and "chemical recovery" during reheating, will be discussed elsewhere. However we note here that, because the chemical ordering times probably conform to the Arrhenius equation, only the slowest chemical ordering processes will avoid becoming diffusion controlled, hence slaved to the structural relaxation, before $T_g$ is reached. The present case is marginal in this respect.

We can relate our observations on chemical order fluctuations to a variety of processes described in the earlier literature. For instance, what we have just described for the cobalt cation probe must also apply to the fluctuations around divalent species such as Ni(II) in glassforming ionic chloride media in which the chloride ions are of high basicity [29,30] due to weak-field counter cations. It will also apply, though with shorter relaxation times, to Ni(II) in chloride media with less basic chloride ions such as glass-forming LiCl-KCl-ZnCl$_2$ solutions [24] and non-glassforming ZnCl$_2$-KCl melts. In the latter case results for the Ni(II) spectra have been reported [31] and the presence of two state (octahedral/tetrahedral) equilibria comparable to the present case have been documented (though no chemical order relaxation times were measured). It appears from these considerations that the chemical order relaxation time problem is the same as the "complex ion lifetime" problem long-debated in molten salt chemistry circles. However some distinction needs to be made between chemical equilibria that change the coordination number, and those that simply involve the change in identity of the caging species at constant coordination number. The former are more likely to couple to the structural relaxation. Both cases can be characterized, in molecular dynamics studies, by the "cage correlation function" [32].

When the cage correlation function that monitors the latter exchanges has a relaxation time longer than the time needed to carry out an electrical transport experiment, negative transport numbers can be observed for the caged cations, and the complex anion concept applies without controversy. However such cases, in which the chemical order relaxation time exceeds the structural relaxation time by more than sixteen orders of magnitude, will be rare, unless we consider species like $CrO_4^{2-}$ in low temperature chromate melts. Such "supercoupling" of anions to cations is the inverse of the 14 orders of magnitude "decoupling" that can be seen in certain "superionic" glasses, at their $T_g$, a parallel to which we will return in our concluding section.

In multicomponent systems with less strongly ordering species, e.g. $Li^+$ in LiCl-KCl solutions, the same chemical ordering phenomenon becomes the "relative mobility" problem in which the mobilities of the cationic species are compared using the chloride ion sublattice as the reference frame [33, 34]. When the chemical ordering is strong, the "weaker" cation has the higher relative mobility.

In the latter simple case it is easily seen that the structural relaxation time scale can be determined by the relative motion of weakly interacting entities (i.e. $nK^+$ with respect to more strongly interacting $LiCl_n^{n-}$ units) while the chemical order relaxation time will be determined by the breakdown of the cage of chloride species around the $Li^+$ cation, as monitored by the cage correlation function for $Li^+$ [35]. Clearly, the stronger the cage binding the longer the cage correlation time at a given temperature. It is here that the parallel between the chemical order problem, thermodynamics vs. kinetics, and the structural order problem, thermodynamics vs. kinetics [17-19] comes into focus, as follows.

The correlation of thermodynamic and kinetic fragilities in single component systems has been discussed in energy landscape topology terms [17-19]. The structure of the energy landscape evidently demands that there be some simple relation between the heights of barriers separating basins and the depths of the basins, because the thermodynamic and kinetic measures of fragility correlate. The landscape for a system with chemical order might be expected to have comparable constraints, possibly dictated by the same factors discussed recently for single component systems by Wales [36] in terms of catastrophy theory. In energy landscape terms the effect of moving along a composition coordinate is dimensionally little different from the effect of moving along a volume coordinate, which we consider in discussing the relation between constant volume and constant pressure measurements [37]. For a fixed composition, or a fixed volume, and a given set of interaction potentials, the system possesses a unique hypersurface whose topology should have common features with those of nearby volumes or compositions, regardless of which variable is being explored. It is to be expected in this case that the systems with strong chemical ordering, according to the non-ideality of mixing indicated by the value of $d^2G/dc^2$, should have long relaxation times for fluctuations involving change of chemical order. And the temperature dependences of the order and of the relaxation times should be correlated. Systems in which the activity coefficients are unity at all compositions, on the other hand, should have chemical order relaxation times indistinguishable from the structural relaxation times.

The difference between single and multicomponent systems is that a deep basin deriving from strong chemical order, in which the cage correlation function is much longer that the structural relaxation time, does not imply that the system is unable to move from minimum to minimum. It can do so, while retaining the "frozen" chemical order, by diffusion of the chemically ordered units (anionic, in the cases we discuss here), rearranging them with respect to the weak-field counter-cations, i.e. the system can be non-ergodic with respect to certain composition fluctuations while remaining freely diffusive. An example would be a system like that of ref. 30 where, at low temperatures, the cage correlation time for the $ZnCl_4^{2-}$ entity would be very long. A more extreme case would be the more fluid $AlCl_4^-$ -containing analog system [29]. In landscape terms this must mean that, within the deep chemical order basin in which such a system is "stuck", there must be substructure within which the system can diffuse and relax by sampling of the substructures, in the same manner as fast ions can freely diffuse in a "superionic" glassy solid electrolytes. Indeed the connection between the two cases is a close one, as we will point out below.

At low temperatures in such a system, the cage correlation time can not only be much longer than any structural relaxation time, but also longer than any experimental time scale, implying broken ergodicity for this degree of freedom, while the system remains freely diffusive. An example is the decrease in glass transition temperature, hence increase in isothermal diffusivity for ions, that is observed [29] when the Lewis acid $FeCl_3$ is added to a molten chloride, forming $FeCl_4^-$ chemically ordered units - which undoubtedly have very long lifetimes near $T_g$.

At intermediate temperatures such frozen chemical order can, in principle become unfrozen on the time scale of experiments and then such phenomena as "melt annealing" [38], which we would prefer to call "chemical aging", can be observed. The classical example is the slow color change in chromium(III) chloride aqueous solutions formed by melting or dissolving one of the various hydrates. More current examples are those in borate melts described by Huaang *et al*. [38] and Khanna [39] in which the slow chemical ordering evidently involves the 3↔4 oxygen coordination of boron. Because of the time scale difference, chemical aging can be observed above the $T_g$.

If the composition of the system is now changed so that the chemically ordered but fast-diffusing sub-systems connect (polymerize), then the diffusivity of the ordered units (hence the structural relaxation time of the chemically ordered system) can change dramatically. Then a glass, which is of course non-ergodic structurally as well as chemically, will form e.g. sodium disilicate glass formed when extra $SiO_2$ is added to sodium orthosilicate $Na_4SiO_4$. However, if the weak-field cations are of small dimensions, then their mobility need not have changed, i.e. they can still move freely at the $T_g$. This is the description of a superionic glass, e.g. $Na_2O.2SiO_2$, with decoupling of the cations from the structure [39]. The system has thus changed from one in which there is supercoupling of the anions to the high field cations within the structurally ordered units, to one in which there is decoupling of the small, weak-field cation subsystem from the chemically ordered polyanionic subsystem.

In summary, we have identified a class of fluctuation, involving chemical order, that can have relaxation times much longer than the structural relaxation time, and whose participants can therefore become frozen (nonergodic on the time scale of the structural relaxation) while the liquid remains above its $T_g$. In extreme cases, in which the chemical order fluctuations are frozen on the time scale of any experiment, the system can act as a single component liquid. When well-decoupled from the structural entropy fluctuations, the chemical order fluctuations appear to be exponential in time and linear in displacement, and probably also Arrhenius in temperature dependence. Slow chemically ordering groups can become kinetically dominant as composition changes cause their polymerization, and can then give rise to superionic glasses. Conditions under which changes in chemical order contribute to the observed liquid "fragility" will be considered elsewhere.

**Methods**

The changes in chemical order were detected using a Cary 14H UV-Vis spectrophotometer with a large sample space into which was placed a aluminium controlled temperature smoothing block within a perspex dry box to avoid problems of moisture condensation on the sample cell. The glassforming mixture of sorbitol, trehalose and tetraethylammonium chloride (46.5:27.8: 25.6 in mole %), $T_{g,cal}$ at 10K/min = 277K, with $3.74 \times 10^{-4}$M $CoCl_2$ was contained in a plastic optical cell (glass cells crack below the glass transition temperature) through which passed a flat copper tube that carried a thermostatted fluid, the temperature of which could be suddenly changed by switching the flow between two independent baths. The effect of temperature on the equilibrium spectrum is shown in Fig. 4. The two state equilibrium evidenced by the isosbestic point (shown in Fig. 4 insert) is less obvious in the present system due to the high symmetry of the octahedral site, and consequent low absorptivity (Laporte-forbidden transition).

Viscosities were measured using a Brookfield digital rotating cylinder viscometer with temperature thermostatted using a Eurotherm controller. The glass transition was determined on hermetically sealed samples using a Perkin-Elmer DSC-7 differential scanning calorimeter at 10K/min.

Acknowledgements:
This work was supported by the NSF under Solid State Chemistry grant no. DMR0082535. We acknowledge helpful discussions with Paul A Madden, and assistance with the experiments from Burkhard Geil and Ranko Richert, see ref. 24, and Marcelo Videa.

**Correspondence and requests for materials should be addressed to C. A. Angell (e-mail: caa@asu.edu).**

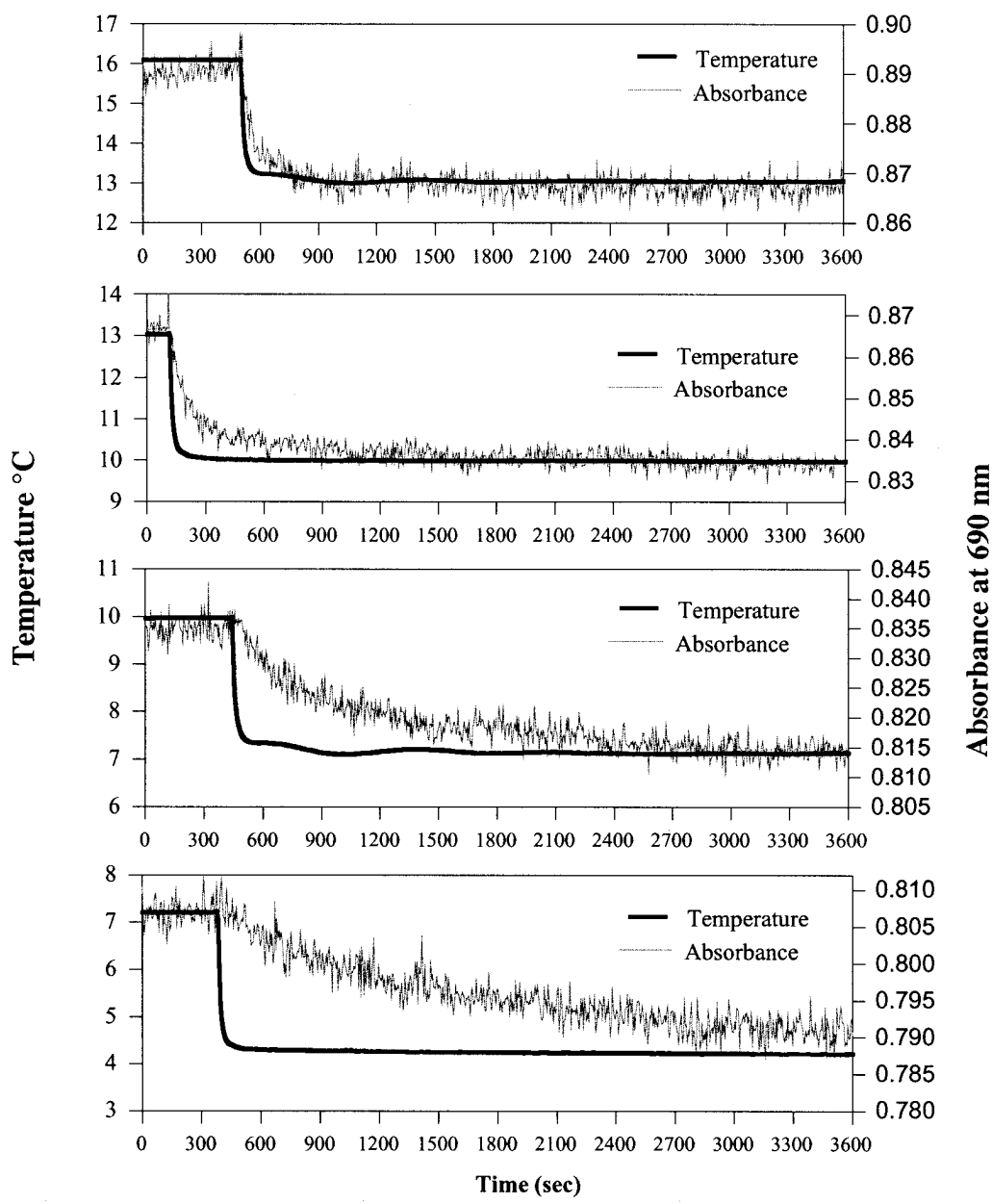

Figure 1. Time evolution of the absorbance at 690nm of the tetrahedral $CoCl_4^{2-}$ site spectrum after 3K T-jumps to different final temperatures, showing the rapidly increasing chemical re-ordering time as temperature decreases. The much faster equilibration of temperature is shown as a solid line.

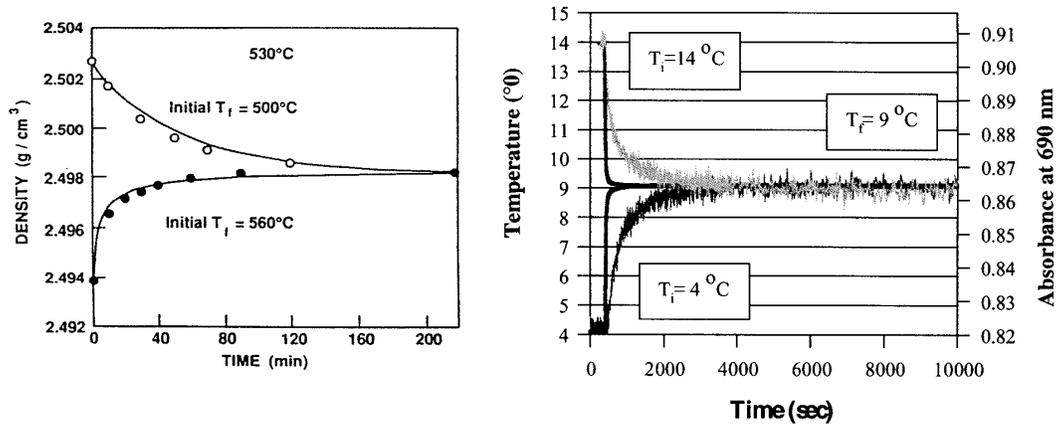

Figure 2. Contrast in the up-jump vs. down jump responses of the present chemical order relaxation (RH panel) and the density relaxation in a classical silicate glass which shows the non-linear response typical of structural relaxation. The equilibration times for the down jumps (lower curve in the silicate glass case) are approximately the same.

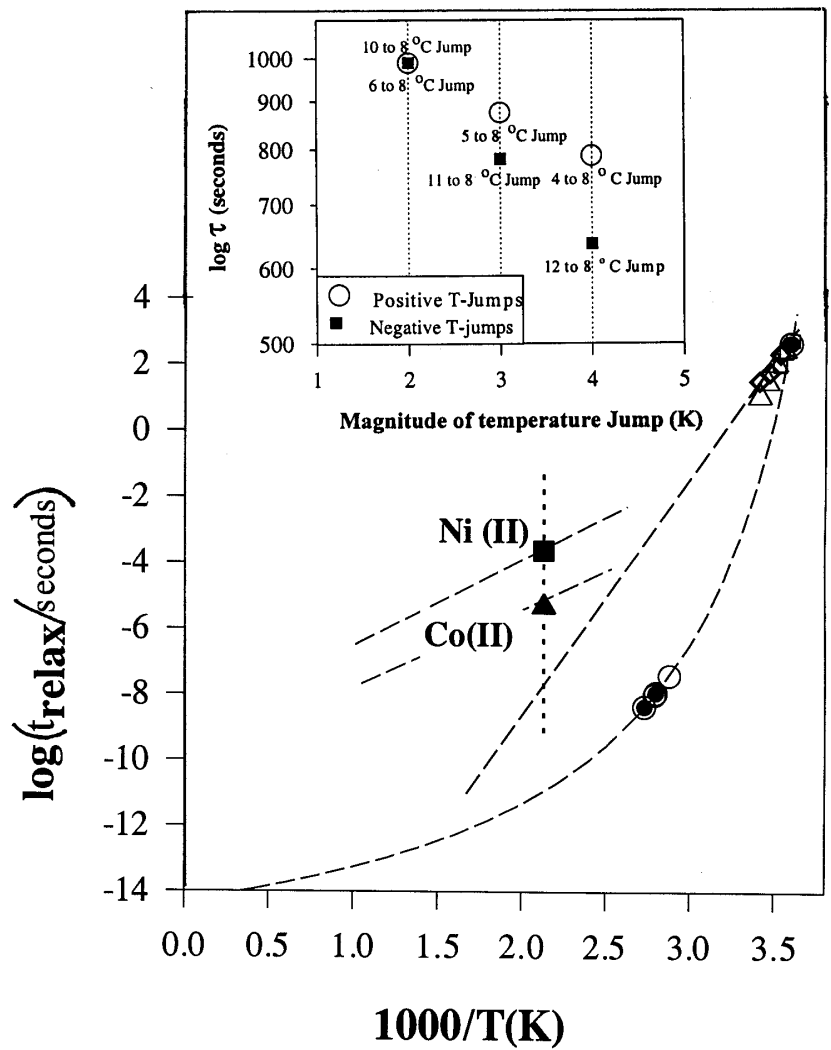

Figure 3. Comparison of the chemical order relaxation time with the shear relaxation time obtained from the viscosity by the Maxwell relation, $\eta = G_\infty \tau$, (and extended to low temperatures using $\eta$ ($T_g$) as an additional point) showing the increasing gap in structural and chemical order relaxation times that opens up as T increases above $T_g$. The points for Co (II) and Ni(II) at higher temperatures are taken from ambient temperature fast T-jump desolvation studies referred to in the text, scaled on to the present system using an isoviscous criterion. They give a crude indication of the values that might be expected for the present system studied on very short time scales. Chemical order relaxation time symbols are for different T-jumps: open squares, 3K, open triangles, 4K: open diamonds, 5K.

(insert). Residual non-linearity in chemical order relaxation demonstrated by dependence of most probable relaxation time, $\tau_0$ of the function $A = A_o \exp[(t/\tau_0)^\beta]$, on the magnitude of the jump for up-jumps vs down jumps.

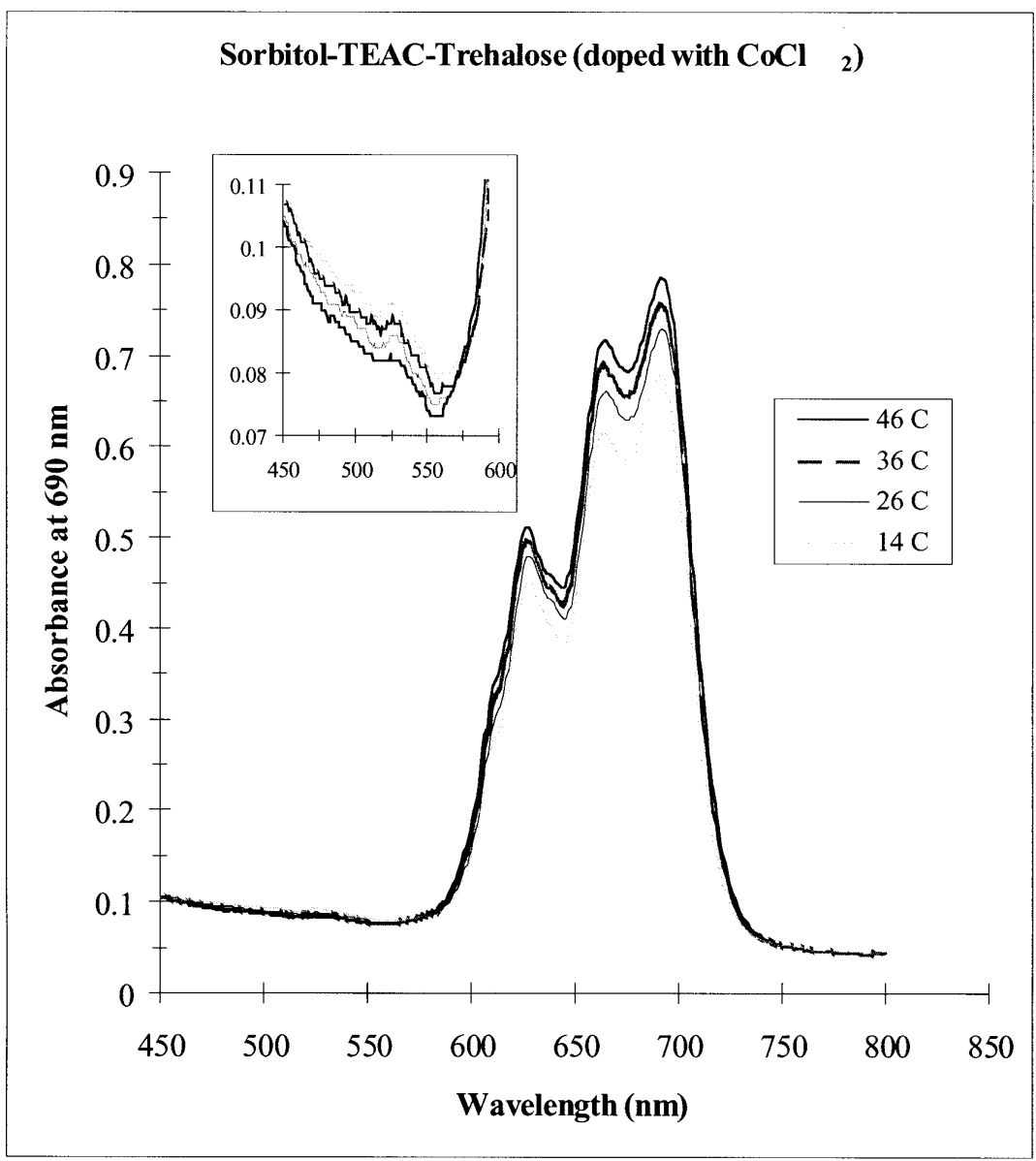

Figure 4. Visible spectrum of Co(II) ( $3.74 \times 10^{-4}$M) in sorbitol-trehalose-tetraethylammonium chloride showing dominance of tetrahedral site absorption, due to strong orbital mixing in the tetrahedral chloride environment. Site population changes following T-jumps were monitored using the 690nm peak.  Insert:  Blowup of 450-600nm region showing weak octohedral Co(II) peak and isosbestic point at ~550nm.